\documentclass[conference]{IEEEtran}
\usepackage{amsmath,amsfonts,amssymb}
\usepackage{algorithmic}
\usepackage{algorithm}
\usepackage{array}
\usepackage{graphicx}
\usepackage{cite}
\usepackage{color}
\usepackage{subcaption}
\usepackage{stfloats}

\graphicspath{{figures/}}
\begin{document}
% =============================================================================
% TITLE
% =============================================================================

\title{Physics-Aware Tensor Reconstruction for Radio Maps in Pixel-Based Fluid Antenna Systems}

\author{\IEEEauthorblockN{Mu Jia\IEEEauthorrefmark{1}\IEEEauthorrefmark{2}\IEEEauthorrefmark{3},
		Hao Sun\IEEEauthorrefmark{4},
		Junting Chen\IEEEauthorrefmark{1}\IEEEauthorrefmark{3} and
		Pooi-Yuen Kam\IEEEauthorrefmark{2}\IEEEauthorrefmark{3}}
	
	\IEEEauthorblockA{\IEEEauthorrefmark{1}School of Science and Engineering,
		\IEEEauthorrefmark{2}School of Artificial Intelligence,\\
		\IEEEauthorrefmark{3}Shenzhen Future Network of Intelligence Institute (FNii-Shenzhen),\\
		The Chinese University of Hong Kong, Shenzhen, Guangdong 518172, China}
	
	\IEEEauthorblockA{\IEEEauthorrefmark{4}Department of Electrical Engineering,
		City University of Hong Kong, Hong Kong}
}

\maketitle   

% =============================================================================
% ABSTRACT
% =============================================================================
\begin{abstract}
	The deployment of pixel-based antennas and fluid antenna systems (FAS) is hindered by prohibitive channel state information (CSI) acquisition overhead. While radio maps enable proactive mode selection, reconstructing high-fidelity maps from sparse measurements is challenging. Existing physics-agnostic or data-driven methods often fail to recover fine-grained shadowing details under extreme sparsity. We propose a Physics-Regularized Low-Rank Tensor Completion (PR-LRTC) framework for radio map reconstruction. By modeling the signal field as a three-way tensor, we integrate environmental low-rankness with deterministic antenna physics. Specifically, we leverage Effective Aerial Degrees-of-Freedom (EADoF) theory to derive a differential gain topology map as a physical prior for regularization. The resulting optimization problem is solved via an efficient Alternating Direction Method of Multipliers (ADMM)-based algorithm. Simulations show that PR-LRTC achieves a 4 dB gain over baselines at a 10\% sampling ratio. It effectively preserves sharp shadowing edges, providing a robust, physics-compliant solution for low-overhead beam management.
\end{abstract}


% =============================================================================
% SECTION I: INTRODUCTION
% =============================================================================

\section{Introduction}
\label{sec:introduction}

The evolution toward Multiple-Input Multiple-Output (MIMO), reconfigurable intelligent surfaces, and fluid antennas is driving the pursuit of unprecedented spectral efficiency \cite{WonShoTon:J21}. However, the high reconfigurability of these systems imposes a prohibitive overhead on CSI acquisition and beam management \cite{KanKim:J25}. To mitigate this, leveraging site-specific a priori information via radio maps—also known as channel knowledge maps (CKMs)—has emerged as a critical enabler \cite{ZenCheXu:J24}. By providing a spatial digital layout of the propagation field, radio maps allow for map-based resource allocation and blind beam prediction, shifting the burden from online signaling to offline environment characterization \cite{SunJiaYu:C25}.

Despite their strategic importance, constructing high-fidelity radio maps is challenged by the high cost of dense spatial sampling in complex environments \cite{liZhaJia:J25}. Conventional interpolation methods or model driven methods often oversmooth shadowing boundaries \cite{TiaCheChe:C24}. While deep learning (DL) approaches offer powerful feature extraction \cite{LevYapKut:J21}, they are inherently data-hungry and struggle with emerging architectures like pixel-based reconfigurable antennas or FAS, where large-scale experimental data are scarce. Furthermore, pure data-driven techniques, such as low-rank tensor completion, remain physics-agnostic, neglecting the underlying electromagnetic propagation laws \cite{SunChe:J24}. Consequently, existing methods struggle to recover fine-grained radiation details from extremely sparse measurements in heterogeneous environments.

In this context, pixel-based fluid antennas emerge not only as a source of complexity but also as a key enabler for high-fidelity reconstruction. While the extensive spatial degrees of freedom in these antennas \cite{ZhaRaoLi:J25,SheWonMur:J26} induce a curse of dimensionality that imposes prohibitive pilot overhead \cite{KiaWonXu:J25}, they simultaneously encode a rigorously deterministic physical structure that has been historically overlooked. Specifically, while absolute signal strength is stochastic, the \textit{differential gain} between antenna modes is physically deterministic and governed by the antenna's electromagnetic structure \cite{zhaRaoMin:J24}. This allows the fluid antenna to act as a physical probe, where mode-dimension coupling provides a rigorous skeleton to regularize the reconstruction process, effectively filling the information gaps left by sparse sampling and data scarcity.

The main contributions of this paper are summarized as follows:
\begin{itemize}
	\item We model the signal field as a three-way tensor to disentangle deterministic antenna physics from stochastic channel fading. Based on the EADoF theory, we derive a differential gain topology map as a rigorous physical prior for reconstruction.
	\item We propose a PR-LRTC framework that fuses sparse measurements with electromagnetic constraints. By introducing a cyclic difference operator and cycle-consistency projection, we design an ADMM-based algorithm that ensures global low-rank consensus and spatial consistency.
	\item Extensive simulations demonstrate that PR-LRTC significantly outperforms conventional and physics-agnostic baselines. It uniquely recovers sharp shadowing edges and fine-grained details under sampling ratios as low as 5\%, overcoming the training data limitations of traditional DL approaches.
\end{itemize}

The remainder of this paper is organized as follows. Section \ref{sec:system_model} describes the system and antenna models. Section \ref{sec:reconstruction} details the PR-LRTC algorithm. Simulation results are presented in Section \ref{sec:simulation}, and Section \ref{sec:conclusion} concludes the paper.

% =============================================================================
% SECTION II: SYSTEM MODEL
% =============================================================================

\section{System Model}
\label{sec:system_model}

\subsection{Geometric Environment and Blocking Model}

\begin{figure}[t] % Changed to [t] to keep it at the top of the column
	\centering
	\includegraphics[width=1\linewidth]{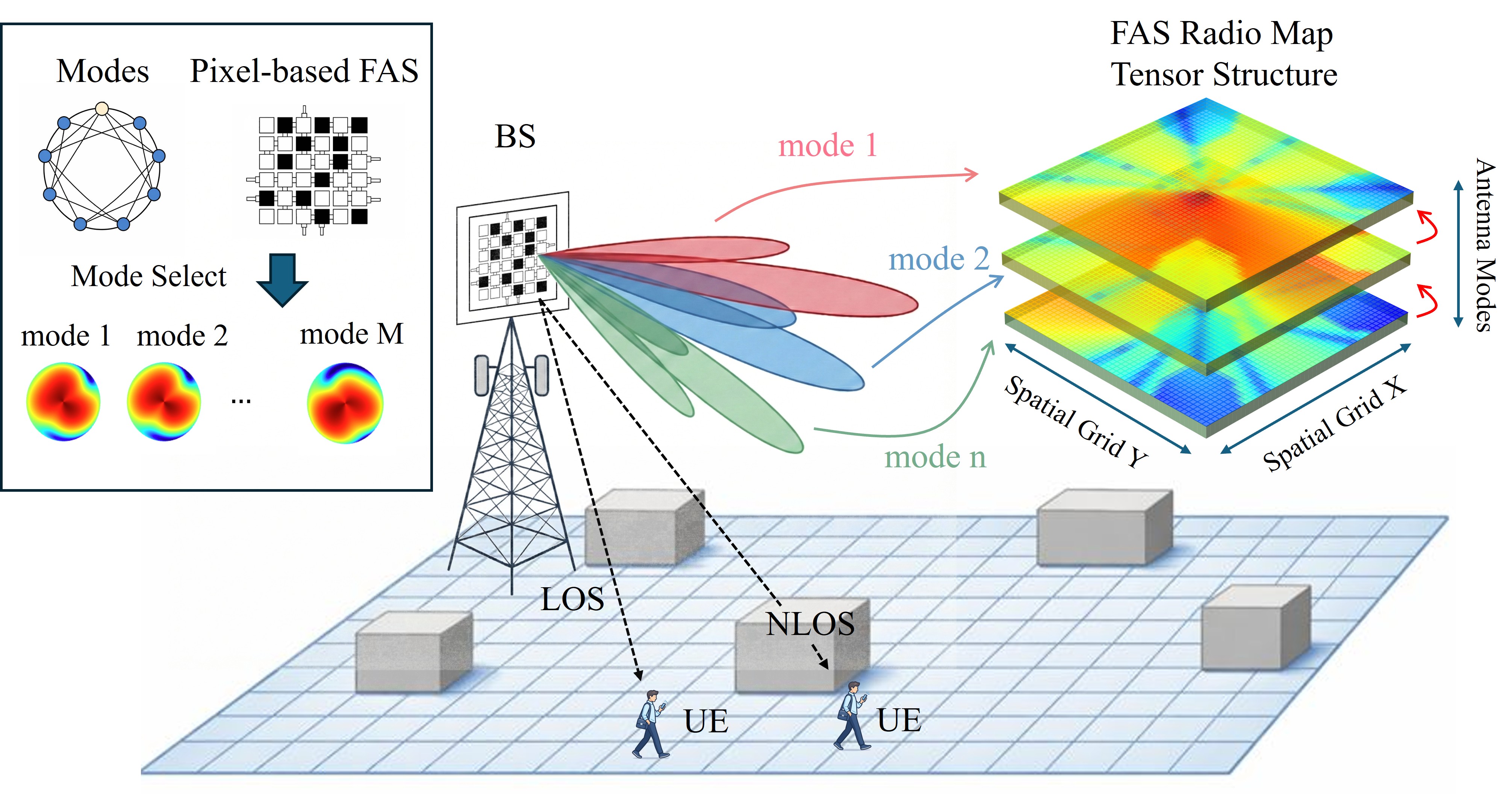} % Check spelling: 'scanerio.jpg' -> 'scenario.jpg'?
	\caption{Pixel-based FAS radio map model. The BS serves users in a cluttered environment. The physical interaction between antenna modes and obstacles induces a structured radio map tensor.}
	\label{Scenario}
\end{figure}

Fig. \ref{Scenario} illustrates the considered downlink communication scenario, bridging the physical propagation environment with the abstract radio map tensor. We consider a downlink wireless communication scenario within a two-dimensional rectangular region $\mathcal{A}$ of dimensions $I \times J$ grids. A single Base Station (BS) equipped with a pixel-based FAS serves single-antenna User Equipments (UEs). The BS is deployed at a fixed coordinate $\mathbf{p}_{\text{BS}} \in \mathbb{R}^2$. The region is populated with a set of obstacles $\{\mathcal{O}_b\}_{b=1}^{N_{\text{obs}}}$, representing physical blockages such as buildings or walls.

To characterize the propagation heterogeneity, we adopt a geometry-based blocking model. For a UE located at $\mathbf{r} \in \mathcal{A}$, the Line-of-Sight (LoS) condition is determined by the geometric visibility between the BS and the UE. Let $\mathcal{L}(\mathbf{p}_{\text{BS}}, \mathbf{r})$ denote the line segment connecting the BS and the UE. We define a binary LoS indicator function $\mathbb{I}_{\text{LoS}}(\mathbf{r}) \in \{0, 1\}$ as:
\begin{equation}
	\label{eq:los_indicator}
	\mathbb{I}_{\text{LoS}}(\mathbf{r}) = 
	\begin{cases} 
		1, & \text{if } \mathcal{L}(\mathbf{p}_{\text{BS}}, \mathbf{r}) \cap \left(\bigcup_{b=1}^{N_{\text{obs}}} \mathcal{O}_b\right) = \varnothing, \\
		0, & \text{otherwise}.
	\end{cases}
\end{equation}
When $\mathbb{I}_{\text{LoS}}(\mathbf{r}) = 1$, the UE is in an LoS region where the direct signal path is unobstructed. Conversely, $\mathbb{I}_{\text{LoS}}(\mathbf{r}) = 0$ indicates a Non-LoS (NLoS) region, where the direct path is blocked, and signal reception relies solely on scattering and reflections.

\subsection{Pixel-Based Reconfigurable Antenna Model}
The BS is equipped with a pixel-based reconfigurable antenna capable of switching among $M$ distinct radiation modes, indexed by $m \in \{1, \dots, M\}$. Each state corresponds to a specific ON/OFF configuration of the RF switches connecting the pixel elements, resulting in a unique surface current distribution.

To capture the physical constraints of the antenna aperture and the structural relationships between different modes, we adopt a physics-aware modeling approach based on the theory of EADoF \cite{ZhaRaoLi:J25}. The complex far-field radiation pattern of the $m$-th mode at an Angle of Departure (AoD) $\phi \in [-\pi, \pi)$, denoted as $G_m(\phi)$, is modeled as a weighted superposition of orthogonal basis functions:
\begin{equation}
	\label{eq:pixel_pattern}
	G_m(\phi) = \sum_{k=-\frac{R-1}{2}}^{\frac{R-1}{2}} w_{m,k} B_k(\phi).
\end{equation}
The components of this model are defined as follows:

\subsubsection{Orthogonal Basis Functions $\{B_k(\phi)\}$}
The radiation capability of any antenna within a finite physical volume is limited by its electromagnetic EADoF, denoted as $R$. We utilize truncated Fourier basis functions to represent the spatial degrees of freedom supported by the antenna aperture:
\begin{equation}
	% Used \jmath to avoid conflict with grid index j
	B_k(\phi) = \frac{1}{\sqrt{2\pi}} e^{j \cdot k \cdot \phi}, \quad k \in \left\{-\frac{R-1}{2}, \dots, \frac{R-1}{2}\right\}.
\end{equation}
For a compact pixel antenna with dimensions around $0.5\lambda$, the EADoF is typically small. We assume $R$ is an odd integer (e.g., $R = 7$) to ensure symmetric spatial frequency coverage, implying that the radiation patterns are composed of low-frequency spatial harmonics.

\subsubsection{Cyclically Correlated Coding Weights $w_{m,k}$}
The switching state $m$ determines the complex weight $w_{m,k}$ for the $k$-th basis mode. In practical reconfigurable antennas, state transitions are often designed to be smooth or cyclical to facilitate efficient beam scanning codebook design. To capture this characteristic, we design the codebook with a circular correlation structure.

Let $\mathbf{w}_m = [w_{m,-\frac{R-1}{2}}, \dots, w_{m,\frac{R-1}{2}}]^T$ denote the weight vector for mode $m$. We generate these weights such that the correlation between any two modes $p$ and $q$ approximately decays with their circular distance in the mode index space:
\begin{equation}
	\label{eq:cyclic_corr}
	\left| \mathbf{w}_p^H \mathbf{w}_q \right| \approx \rho^{\delta_{\text{circ}}(p, q)}
\end{equation}
where $\rho \in [0, 1]$ is the correlation coefficient reflecting the similarity between adjacent states, and $\delta_{\text{circ}}(p, q) = \min(|p-q|, M-|p-q|)
$ is the circular distance.

\subsection{Received Signal Model}
\label{sec:signal_model}

Based on the explicit blockage geometry and the reconfigurable radiation characteristics defined above, we formulate the received signal power at a UE located at a macroscopic coordinate $\mathbf{r} = [x, y]^T$. Let $P_{\text{tx}}$ denote the transmit power in dBm. The Received Signal Strength (RSS) is governed by the interplay between the deterministic antenna gain, the geometry-dependent path loss, and the stochastic shadowing effects. Due to the presence of physical obstacles, the propagation law exhibits a regime-switching behavior conditioned on the visibility state $\mathbb{I}_{\text{LoS}}(\mathbf{r})$.

We first define the geometric features of the communication link. The Euclidean distance between the BS and the UE is denoted as $d(\mathbf{r}) = \|\mathbf{r} - \mathbf{p}_{\text{BS}}\|_2$, and the geometric AoD $\phi(\mathbf{r})$ is given by $\phi(\mathbf{r}) = \operatorname{atan2}(y - y_{\text{BS}}, x - x_{\text{BS}})$. Unlike conventional systems with static sectors, the antenna gain in our system is coupled with the switching mode $m$. Substituting the AoD into the basis expansion model \eqref{eq:pixel_pattern}, the logarithmic directional gain $\mathcal{G}(\mathbf{r}, m)$ (in dBi) for the $m$-th mode is derived as:
\begin{equation}
	\label{eq:gain_model}
	\mathcal{G}(\mathbf{r}, m) = 10 \log_{10} \left( \left| \sum_{k=-\frac{R-1}{2}}^{\frac{R-1}{2}} w_{m,k} B_k(\phi(\mathbf{r})) \right|^2 \right).
\end{equation}
This term encapsulates the reconfigurability of the pixel antenna, acting as a controllable spatial filter that weights the signal power based on the chosen current distribution $\mathbf{w}_m$.

To capture the spatial heterogeneity caused by blockages, we adopt a segmented log-distance path loss model \cite{SunJiaYu:C25}. The received power $P_r(\mathbf{r}, m)$ is modeled as a superposition of the transmit power, the mode-dependent antenna gain, the path loss, and a shadowing term. Specifically, the propagation parameters switch between the LoS region and the NLoS region. The RSS model is expressed as: 
\begin{equation}
	\label{eq:rss_model}
	P_r(\mathbf{r}, m) = P_{\text{tx}} + \mathcal{G}(\mathbf{r}, m) - \left( \alpha_{\kappa} 10 \log_{10}(d(\mathbf{r})) + \beta_{\kappa} \right) + \varepsilon_{\kappa},
\end{equation}
where the subscript $\kappa \in \{\text{L}, \text{N}\}$ indicates the propagation state determined by the indicator function $\mathbb{I}_{\text{LoS}}(\mathbf{r})$. Here, $\alpha_{\kappa}$ represents the path loss exponent, and $\beta_{\kappa}$ accounts for the reference attenuation at 1 meter. The term $\varepsilon_{\kappa} \sim \mathcal{N}(0, \sigma_{\kappa}^2)$ represents the log-normal shadowing effect, which captures the random signal fluctuations due to scattering and diffraction not modeled by the deterministic path loss.

\subsection{Radio Map Representation and Physical Priors}
\label{sec:radio_map}

To facilitate the reconstruction of the radio environment, we discretize the continuous region $\mathcal{A}$ into a uniform grid of $I \times J$ pixels. Let $\mathbf{r}_{i,j}$ denote the coordinate of the $(i,j)$-th grid element. The objective is to characterize the signal field across all possible antenna configurations.

\subsubsection{The Global RSS Tensor ($\mathcal{X}$)}
We define the Pixel-Mode Radio Map as a 3-way tensor $\mathcal{X} \in \mathbb{R}^{I \times J \times M}$. Each entry $\mathcal{X}_{i,j,m}$ represents the long-term stable RSS at location $\mathbf{r}_{i,j}$ when the BS operates in mode $m$:
\begin{equation}
	\label{eq:rss_tensor}
	\mathcal{X}_{i,j,m} = P_r(\mathbf{r}_{i,j}, m).
\end{equation}
According to the model in \eqref{eq:rss_model}, $\mathcal{X}$ encapsulates three levels of information: macroscopic path loss (spatial trend), local shadowing (environmental geometry), and mode-dependent gains (antenna physics). This tensor serves as the ground truth for our reconstruction problem. Due to the limited EADoF of the antenna and the spatial correlation of the shadowing field, $\mathcal{X}$ naturally exhibits a low-rank structure.

\subsubsection{The Physics-Aware Differential Prior ($\mathcal{D}$)}
A major challenge in radio map completion is that the RSS values are heavily corrupted by unpredictable shadowing $\varepsilon_{\kappa}$ and environmental blockages. However, the pixel antenna structure provides a deterministic coupling between different modes. We define a Differential Gain Topology Map $\mathcal{D} \in \mathbb{R}^{I \times J \times M \times M}$ to capture this physical invariant.

For any two modes $p$ and $q$ at location $\mathbf{r}_{i,j}$, the expected RSS difference is derived from \eqref{eq:rss_model} as:
\begin{equation}
	\label{eq:diff_gain_map}
	[\mathcal{D}_{i,j}]_{p, q} = \mathcal{G}(\mathbf{r}_{i,j}, p) - \mathcal{G}(\mathbf{r}_{i,j}, q),
\end{equation}
Note that because the path loss and shadowing terms are mode-independent, they cancel out in the subtraction. This leads to two critical observations that bridge the system model and the reconstruction algorithm:
\begin{itemize}
	\item \textit{Environmental Robustness:} Unlike the absolute RSS $\mathcal{X}_{i,j,m}$, the differential map $\mathcal{D}$ is independent of specific blockage instances and shadowing variance. It depends solely on the antenna's basis weights $\mathbf{w}_m$ and the geometric AoD $\phi(\mathbf{r}_{i,j})$.
	\item \textit{Structural Regularization:} The map $\mathcal{D}$ provides a soft physical skeleton. For any reconstructed tensor $\widehat{\mathcal{X}}$, its internal variations along the mode dimension should ideally align with the theoretical gradients provided by $\mathcal{D}$, i.e., $\widehat{\mathcal{X}}_{i,j,p} - \widehat{\mathcal{X}}_{i,j,q} \approx [\mathcal{D}_{i,j}]_{p, q}$.
\end{itemize}

% =============================================================================
% SECTION III: PROPOSED FRAMEWORK
% =============================================================================
\section{Physics-Regularized Tensor Reconstruction}
\label{sec:reconstruction}

In this section, we formulate the reconstruction of the global RSS tensor $\mathcal{X} \in \mathbb{R}^{I \times J \times M}$ as a physics-constrained inverse problem. We propose a PR-LRTC framework that fuses sparse measurements with the electromagnetic priors derived in Section~\ref{sec:system_model}.

\subsection{Problem Formulation}
Let $\Omega \subset \{1,\dots,I\} \times \{1,\dots,J\} \times \{1,\dots,M\}$ denote the set of indices corresponding to available measurements. We introduce the sampling operator $\mathcal{P}_{\Omega}(\cdot)$, which retains the tensor entries indexed by $\Omega$ and sets the remaining entries to zero. The observed sparse tensor $\mathcal{Y}$ is modeled as a noisy, incomplete version of the ground truth $\mathcal{X}$:
\begin{align}
	\label{eq:meas_model}
	\mathcal{Y} = \mathcal{P}_{\Omega}(\mathcal{X} + \mathcal{N}),
\end{align}
where $\mathcal{N}$ represents additive measurement noise. This formulation implies that $\mathcal{Y}_{i,j,m} = 0$ for all $(i,j,m) \notin \Omega$.

To recover $\mathcal{X}$ from the sparse observation $\mathcal{Y}$, we formulate an inverse problem regularized by two structural priors:
\begin{enumerate}
	\item \textit{Global Multilinear Low-Rankness:} The tensor $\mathcal{X}$ exhibits low-rank structure across its unfoldings, induced by spatial shadowing correlations and limited antenna EADoF.
	\item \textit{Local Differential Consistency:} The inter-mode RSS variations at each location must align with the deterministic differential topology $\mathcal{D}$.
\end{enumerate}

Accordingly, we reconstruct $\mathcal{X}$ by solving the following convex optimization problem:
\begin{align}
	\label{eq:optimization_problem}
	\min_{\mathcal{X}} \quad & \frac{1}{2} \big\| \mathcal{P}_{\Omega}(\mathcal{X} - \mathcal{Y}) \big\|_F^2 + \lambda_2 \mathcal{R}_{\text{phys}}(\mathcal{X}, \mathcal{D}) \notag \\
	& + \lambda_1 \sum_{k=1}^{3} \alpha_k \big\| \mathbf{X}_{(k)} \big\|_*,
\end{align}
where the first term enforces data fidelity on the observed set $\Omega$. The second term $\mathcal{R}_{\text{phys}}(\mathcal{X}, \mathcal{D})$ by the differential gain topology $\mathcal{D}$ (to be explicitly formulated in Section III-B). The term $\|\mathbf{X}_{(k)}\|_*$ denotes the nuclear norm of the mode-$k$ unfolding of $\mathcal{X}$, defined as the sum of its singular values. The weighted sum $\sum_{k=1}^{3} \alpha_k \|\mathbf{X}_{(k)}\|_*$ constitutes the \textit{overlapped nuclear norm}, which serves as a tight convex surrogate for the multilinear tensor rank. The regularization parameters $\lambda_1, \lambda_2 > 0$ balance the trade-off between global low-rank structure and physical consistency, while $\{\alpha_k\}_{k=1}^3$ are non-negative weights satisfying $\sum_k \alpha_k = 1$.

\subsection{Physics-Aware Regularizer Design}
The core of our regularization strategy lies in a compact mathematical formulation of the circular differential constraints induced by the pixel antenna's mode topology. For a specific spatial location $(i,j)$, let $\mathbf{x}_{i,j} = \mathcal{X}_{i,j,:} \in \mathbb{R}^{M}$ denote the mode-RSS vector. We define the \textit{theoretical differential vector} $\mathbf{d}_{i,j} \in \mathbb{R}^M$ derived from the map $\mathcal{D}$ as:
\begin{align}
	\label{eq:vector_defs}
	[\mathbf{d}_{i,j}]_m &= [\mathcal{D}_{i,j}]_{m,\, m \ominus 1}, \quad m=1,\dots,M,
\end{align}
where the operator $\ominus$ denotes circular subtraction over the index set $\{1,\dots,M\}$.

\subsubsection*{Cycle-Consistency Projection}
In a closed loop topology, the algebraic sum of differences between adjacent nodes must theoretically vanish. However, the raw differential estimates from $\mathcal{D}$ may contain accumulated modeling errors. To enforce geometric validity, we project $\mathbf{d}_{i,j}$ onto the cycle-consistent subspace by removing its mean component:
\begin{align}
	\label{eq:dij_projection}
	\tilde{\mathbf{d}}_{i,j} = \left( \mathbf{I} - \frac{1}{M}\mathbf{1}\mathbf{1}^T \right) \mathbf{d}_{i,j},
\end{align}
where $\mathbf{I}$ is the identity matrix and $\mathbf{1} \in \mathbb{R}^M$ is the all-ones vector. In the sequel, we refer to the $\tilde{\mathbf{d}}_{i,j}$ simply as $\mathbf{d}_{i,j}$.

\subsubsection*{Cyclic Difference Operator}
We introduce the cyclic first-order difference operator $\mathbf{A} \in \mathbb{R}^{M \times M}$, which acts as the incidence matrix of the mode cycle graph. Its action on the state vector is defined as $(\mathbf{A}\mathbf{x})_m = x_m - x_{m\ominus 1}$. The explicit matrix representation is given by:
\begin{align}
	\label{eq:A_matrix}
	\mathbf{A} = 
	\begin{bmatrix}
		1 & 0 & \cdots & 0 & -1 \\
		-1 & 1 & \cdots & 0 & 0 \\
		\vdots & \ddots & \ddots & \vdots & \vdots \\
		0 & 0 & \cdots & -1 & 1
	\end{bmatrix}.
\end{align}
Consequently, we define the physics-aware regularizer as the aggregate squared Euclidean deviation between the reconstructed gradients and the theoretical differentials:
\begin{align}
	\label{eq:phys_regularizer_matrix}
	\mathcal{R}_{\text{phys}}(\mathcal{X}, \mathcal{D}) = \sum_{i=1}^{I}\sum_{j=1}^{J} \big\| \mathbf{A}\mathbf{x}_{i,j} - \mathbf{d}_{i,j} \big\|_2^2.
\end{align}
This term functions as a soft constraint, robustly propagating information across the mode dimension while accommodating minor discrepancies between the idealized antenna model and practical measurements.

\subsection{Optimization via ADMM}
The optimization problem in \eqref{eq:optimization_problem} is convex; however, the simultaneous minimization of multiple nuclear norms involving different tensor unfoldings renders the problem non-smooth and interdependent. Direct gradient-based methods are essentially inapplicable here. To address this, we employ the ADMM framework, which allows us to decompose the complex global problem into smaller, tractable sub-problems via \textit{variable splitting}.

We introduce auxiliary tensor variables $\{\mathcal{M}_k\}_{k=1}^3$ to decouple the overlapped nuclear norms. This transforms the unconstrained problem into a constrained one:
\begin{align}
	\label{eq:admm_form}
	\min_{\mathcal{X},\{\mathcal{M}_k\}} \quad & \frac{1}{2}\big\| \mathcal{P}_{\Omega}(\mathcal{X} - \mathcal{Y}) \big\|_F^2 + \lambda_2 \mathcal{R}_{\text{phys}}(\mathcal{X}, \mathcal{D}) \notag \\
	& + \lambda_1 \sum_{k=1}^{3} \alpha_k \big\| \mathbf{M}_{k(k)} \big\|_* \\
	\text{s.t.} \quad & \mathcal{X} = \mathcal{M}_k, \quad k=1,2,3, \notag
\end{align}
where $\mathbf{M}_{k(k)}$ denotes the mode-$k$ unfolding of $\mathcal{M}_k$. This formulation isolates the non-smooth nuclear norm terms onto the auxiliary variables $\mathcal{M}_k$, leaving $\mathcal{X}$ to handle the smooth data fidelity and physical regularization terms.

The augmented Lagrangian function, incorporating scaled dual variables $\{\mathcal{U}_k\}_{k=1}^3$ and a penalty parameter $\rho > 0$, is expressed as:
\begin{align}
	\label{eq:aug_lag}
	\mathcal{L}_{\rho}&(\mathcal{X},\{\mathcal{M}_k\},\{\mathcal{U}_k\}) = \frac{1}{2}\big\| \mathcal{P}_{\Omega}(\mathcal{X} - \mathcal{Y}) \big\|_F^2 \notag \\
	&+ \lambda_2 \sum_{i,j} \big\| \mathbf{A}\mathbf{x}_{i,j} - \mathbf{d}_{i,j} \big\|_2^2 + \sum_{k=1}^{3} \lambda_1 \alpha_k \big\| \mathbf{M}_{k(k)} \big\|_* \notag \\
	&+ \sum_{k=1}^{3} \frac{\rho}{2} \left( \big\| \mathcal{X} - \mathcal{M}_k + \mathcal{U}_k \big\|_F^2 - \|\mathcal{U}_k\|_F^2 \right).
\end{align}
The ADMM algorithm proceeds by minimizing $\mathcal{L}_{\rho}$ with respect to each block of variables alternately.

\subsubsection{\textbf{Primal Update $\mathcal{X}$ (Physics-Data Fusion)}}
The minimization w.r.t. $\mathcal{X}$ acts as a \textit{fusion step} that integrates information from sparse measurements, physical laws, and the low-rank consensus. Since the objective function involves only quadratic terms concerning $\mathcal{X}$, this step admits a closed-form solution. Furthermore, the problem naturally decouples across the spatial domain, allowing for parallel computation of $I \times J$ independent sub-problems.

For a specific location $(i,j)$, let $\mathbf{y}_{i,j}$ be the zero-filled measurement vector and $\mathbf{P}_{i,j}$ be the diagonal observation mask. Setting the gradient $\nabla_{\mathbf{x}_{i,j}} \mathcal{L}_{\rho} = 0$ leads to the following linear system:
\begin{align}
	\label{eq:linear_system}
	\Big(\underbrace{\mathbf{P}_{i,j}}_{\text{Data}} &+ \underbrace{2\lambda_2 \mathbf{A}^T\mathbf{A}}_{\text{Physics}} + \underbrace{3\rho\mathbf{I}}_{\text{Consensus}}\Big)\mathbf{x}_{i,j} \notag \\
	&= \mathbf{P}_{i,j}\mathbf{y}_{i,j} + 2\lambda_2 \mathbf{A}^T\mathbf{d}_{i,j} + \rho \sum_{k=1}^{3} (\mathcal{M}_k - \mathcal{U}_k)_{i,j}.
\end{align}
The coefficient matrix in \eqref{eq:linear_system} represents a regularized Hessian, where $\mathbf{L}_{\text{cyc}} = \mathbf{A}^T\mathbf{A}$ is the Laplacian of the mode cycle graph enforcing circular smoothness. Given that the number of antenna modes is typically small in practical pixel antenna systems, the dimensionality of this linear system is low. Consequently, we solve \eqref{eq:linear_system} efficiently via direct matrix inversion (e.g., Cholesky decomposition or LU factorization), which is computationally negligible compared to the SVD steps in the auxiliary update.

\subsubsection{\textbf{Auxiliary Update $\mathcal{M}_k$ (Low-Rank Projection)}}
The optimization w.r.t. $\mathcal{M}_k$ aims to find a tensor that is close to the current estimate $\mathcal{X}$ (adjusted by the dual variable) while minimizing the nuclear norm rank proxy. Mathematically, this corresponds to the \textit{proximal operator} of the nuclear norm:
\begin{align}
	\mathcal{M}_k^{t+1} = \arg \min_{\mathcal{M}_k} \frac{\rho}{2} \big\| \mathcal{M}_k - (\mathcal{X}^{t+1} + \mathcal{U}_k^t) \big\|_F^2 + \lambda_1 \alpha_k \big\| \mathbf{M}_{k(k)} \big\|_*.
\end{align}
The closed-form solution is given by the Singular Value Thresholding (SVT) operator:
\begin{align}
	\label{eq:svt_update}
	\mathbf{M}_{k(k)}^{t+1} = \mathcal{D}_{\tau_k} \left( \mathbf{X}_{(k)}^{t+1} + \mathbf{U}_{k(k)}^{t} \right), \quad \tau_k = \frac{\lambda_1 \alpha_k}{\rho},
\end{align}
where $\mathcal{D}_{\tau}(\mathbf{Z})$ applies soft-thresholding to the singular values of matrix $\mathbf{Z}$. This step explicitly enforces the global environmental consistency by suppressing noise-induced high-rank components.

\subsubsection{\textbf{Dual Update}}
The dual variables are updated via standard gradient ascent to enforce the equality constraints asymptotically:
\begin{align}
	\label{eq:dual_update}
	\mathcal{U}_k^{t+1} = \mathcal{U}_k^{t} + \left( \mathcal{X}^{t+1} - \mathcal{M}_k^{t+1} \right).
\end{align}

\subsubsection{\textbf{Convergence Criterion}}
The algorithm terminates when the primal residual $r^{(t)}$ and dual residual $s^{(t)}$ fall below predefined tolerances $\epsilon_{\text{pri}}$ and $\epsilon_{\text{dual}}$, respectively:
\begin{align}
	\label{eq:stopping}
	r^{(t)} &= \max_{k} \|\mathcal{X}^{(t)} - \mathcal{M}_k^{(t)}\|_F \le \epsilon_{\text{pri}}, \notag \\
	s^{(t)} &= \rho \max_{k} \|\mathcal{M}_k^{(t)} - \mathcal{M}_k^{(t-1)}\|_F \le \epsilon_{\text{dual}}.
\end{align}
The overall computational complexity is dominated by the SVD in the $\mathcal{M}_k$-update, scaling with $\mathcal{O}(IJM \cdot \min(IJ, M))$. Since the mode dimension $M$ is typically small (e.g., $M =12$) in pixel antenna systems, the reconstruction is computationally efficient.
The complete reconstruction procedure is summarized in Algorithm \ref{alg:pr_lrtc}.

\begin{algorithm}[htbp]
	\caption{PR-LRTC for Pixel Antenna Radio Map}
	\label{alg:pr_lrtc}
	\begin{algorithmic}[1]
		\STATE \textbf{Input:} Sparse observations $\mathcal{Y}_{\Omega}$, Differential map $\mathcal{D}$, parameters $\lambda_1, \lambda_2, \rho$, weights $\{\alpha_k\}$.
		\STATE \textbf{Preprocessing:} Compute cyclic difference matrix $\mathbf{A}$ and project $\mathbf{d}_{i,j}$ via \eqref{eq:dij_projection}.
		\STATE \textbf{Initialize:} $\mathcal{X}^0, \{\mathcal{M}_k^0\}, \{\mathcal{U}_k^0\} \leftarrow \mathbf{0}$.
		\WHILE{convergence criteria \eqref{eq:stopping} not met}
		\FOR{each location $(i,j)$ in parallel}
		\STATE Construct RHS and system matrix for \eqref{eq:linear_system}.
		\STATE Solve for $\mathbf{x}_{i,j}$.
		\STATE Update $\mathcal{X}_{i,j,:} \leftarrow \mathbf{x}_{i,j}$.
		\ENDFOR
		\FOR{$k=1$ to $3$}
		\STATE Unfold $\mathcal{X}$ to $\mathbf{X}_{(k)}$; Unfold $\mathcal{U}_k$ to $\mathbf{U}_{k(k)}$.
		\STATE $\mathbf{M}_{k(k)} \leftarrow \text{SVT}(\mathbf{X}_{(k)} + \mathbf{U}_{k(k)}, \lambda_1\alpha_k/\rho)$.
		\STATE Refold $\mathbf{M}_{k(k)}$ to $\mathcal{M}_k$.
		\STATE Update Dual: $\mathcal{U}_k \leftarrow \mathcal{U}_k + (\mathcal{X} - \mathcal{M}_k)$.
		\ENDFOR
		\ENDWHILE
		\STATE \textbf{Output:} Reconstructed Tensor $\widehat{\mathcal{X}}$.
	\end{algorithmic}
\end{algorithm}

\begin{figure*}[htbp]
	\centering
	% --- 第一行：方法对比 (结果最重要，放上面) ---
	\begin{subfigure}{1\linewidth}
		\centering
		% 建议适当调整 trim 或 clip 去掉图片白边，或者保持现状
		\includegraphics[width=\linewidth]{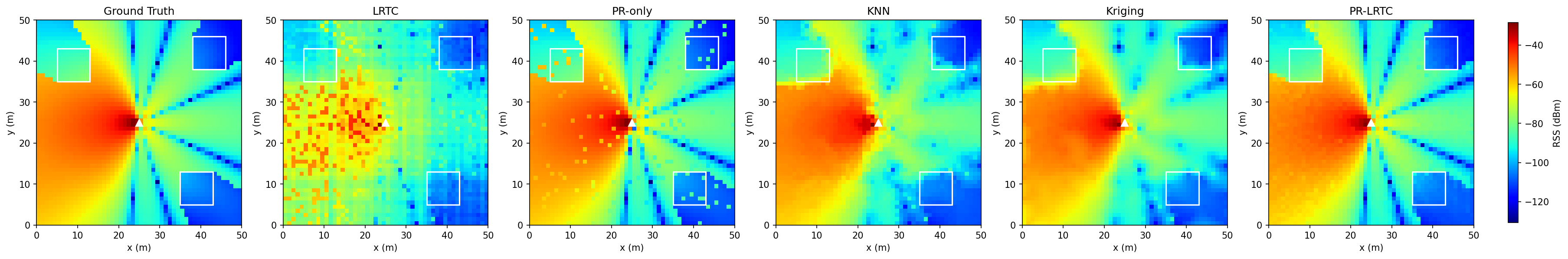}
		\caption{Comparison of reconstructed RSS maps (Mode 0).}
		\label{fig:sub_comparison}
	\end{subfigure}
	
	\vspace{2pt} % 两行图片之间的间距
	
	% --- 第二行：3D 张量展示 (作为补充视角) ---
	\begin{subfigure}{1\linewidth}
		\centering
		\includegraphics[width=\linewidth]{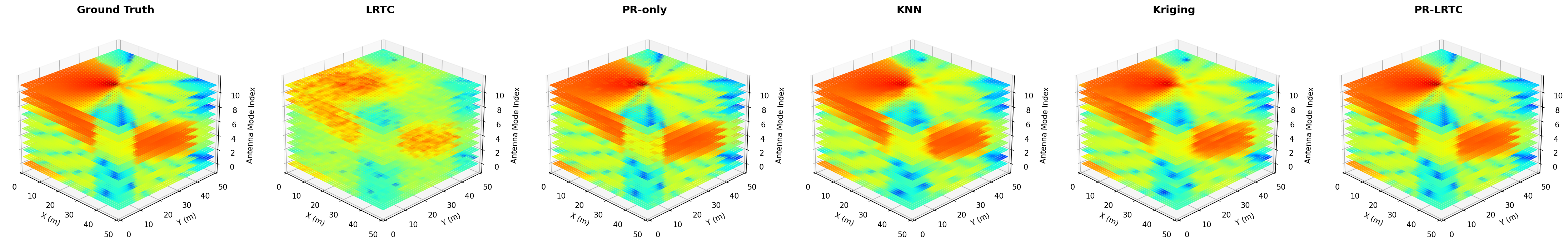}
		\caption{Ground Truth 3D Radio Map Tensor across different antenna modes.}
		\label{fig:sub_tensor}
	\end{subfigure}
	
	\caption{Comparison of reconstruction results at 10\% sampling ratio. (a) compares the reconstruction details of Mode 0, where the proposed PR-LRTC best preserves the shadowing edges. (b) illustrates the complex 3D structure of the radio map tensor.}
	\label{fig:combined_view}
\end{figure*}

% =============================================================================
% SECTION IV: SIMULATION RESULTS
% =============================================================================
\section{Simulation Results}
\label{sec:simulation}

\subsection{Simulation Setup}
\label{subsec:sim_setup}

We simulate a downlink scenario in a $50 \times 50$ m$^2$ area discretized into a $50 \times 50$ grid ($I=J=50$). The BS is located at the center. The environment contains three obstacles of size $8 \times 8$ m, creating distinct LoS and NLoS regions.

\subsubsection{Antenna and Channel Parameters}
The BS is equipped with a pixel-based reconfigurable antenna operating with $M=12$ modes. The radiation patterns are generated based on the EADoF theory with $R=7$, ensuring a maximum gain of approximately 7.14 dBi. To model practical hardware constraints, the codebook is designed with a high cyclic correlation coefficient $\rho = 0.96$.

The channel follows the 3GPP TR 38.901 UMi model \cite{ZhuWanHua:B21}. The path loss exponents are set to $\alpha_{\text{L}} = 2.0$ for LoS and $\alpha_{\text{N}} = 3.8$ for NLoS. The shadowing standard deviation is $\sigma_{\text{L}} = 1.0$ dB and $\sigma_{\text{N}} = 3.0$ dB, respectively. The transmit power is set to $P_{\text{tx}} = 30$ dBm.

\subsubsection{Algorithm Hyperparameters}
The ground truth tensor $\mathcal{X}$ is constructed using \eqref{eq:rss_tensor}. For the PR-LRTC algorithm, the weights for the overlapped nuclear norm are set to $\alpha_k = 1/3$. The regularization parameters are tuned to $\lambda_1 = 0.1$ (low-rank weight) and $\lambda_2 = 5.0$ (physics weight), with an ADMM penalty parameter $\rho = 1.0$.

% [SUGGESTION]: Combine Fig 1 (Env) and Fig 2 (Patterns) into a single sub-figure or side-by-side figure here.
%\begin{figure}[t]
%	\centering
%	\includegraphics[width=0.7\linewidth]{fig1_environment.png}
%	\caption{Simulation environment layout.}
%	\label{fig:setup}
%\end{figure}

\subsection{Performance Analysis}
We evaluate the proposed framework against four baselines: 
1) \textit{KNN}, a spatial interpolation method based on inverse distance weighting that treats antenna modes independently; 
2) \textit{Kriging}, a geostatistical interpolation method based on Gaussian process regression that utilizes spatial covariance for estimation;
3) \textit{LRTC} (Standard Low-Rank Tensor Completion), which corresponds to our model with $\lambda_2=0$, relying solely on global data correlation without physical priors; 
and 4) \textit{PR-only} (Physics-Regularized Reconstruction), an ablation variant with $\lambda_1=0$ that enforces differential gain consistency but neglects the global low-rank structure.

\begin{figure}[htbp]
	\centering
	\includegraphics[width=1\linewidth]{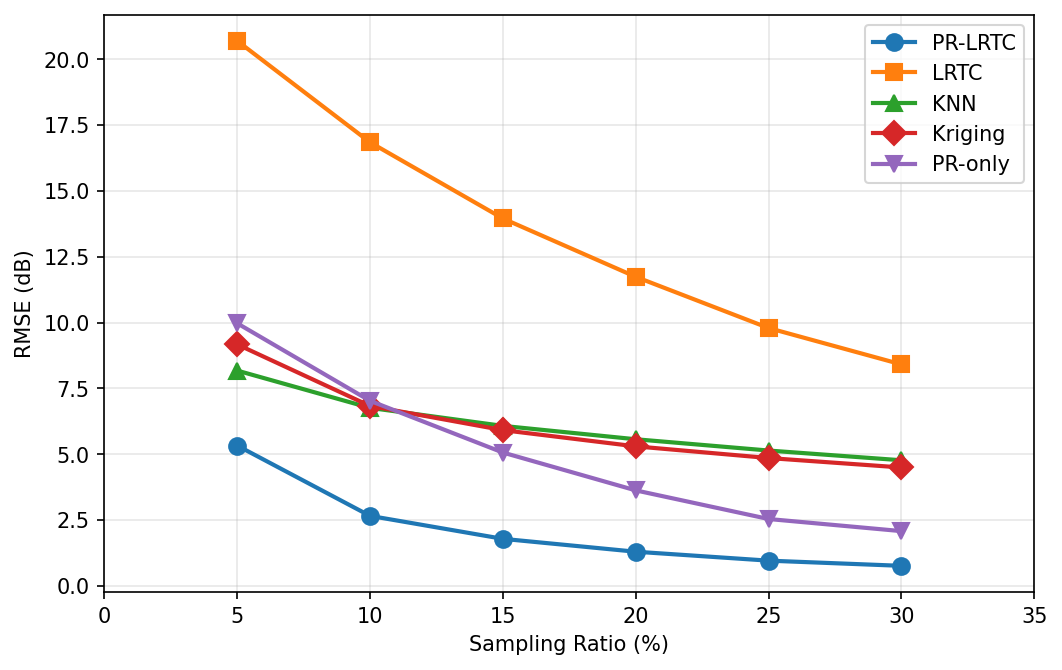}
	\caption{RMSE (dB) comparison versus sampling ratio.}
	\label{fig:nmse_performance}
\end{figure}

Fig. \ref{fig:nmse_performance} presents the reconstruction RMSE across varying sampling ratios, where the proposed PR-LRTC consistently achieves the best performance. The reasons behind these performance gaps are clearly reflected in Fig. \ref{fig:combined_view}. Specifically, the standard LRTC focuses solely on the low-rank structure across modes while neglecting geospatial continuity, resulting in spatial discontinuity. Conversely, the PR-only method relies on geographical features but ignores the intrinsic correlations between different antenna modes. Furthermore, conventional spatial interpolation methods, such as \textit{KNN} and \textit{Kriging}, produce overly smooth estimates where the deep fading regions (indicated by the dark blue zones) are broken or blurred. While Kriging slightly outperforms KNN by exploiting spatial covariance, both fail to capture the sharp shadowing edges caused by blockages. In contrast, our PR-LRTC effectively integrates both spatial constraints and mode correlations, thereby successfully recovering these fine-grained details and achieving superior reconstruction fidelity.

% =============================================================================
% SECTION V: CONCLUSION
% =============================================================================
\section{Conclusion}
\label{sec:conclusion}
This paper proposed a physics-regularized tensor reconstruction framework for high-fidelity radio maps in pixel-based FAS. By modeling the multi-mode radio map as a three-way tensor, we integrated environmental low-rankness with deterministic antenna physics via an EADoF-based differential gain prior. The proposed ADMM-based algorithm effectively resolves mode-dependent gain patterns amidst stochastic environmental noise. Simulations demonstrate that our framework reliably recovers fine-grained signal structures and sharp shadowing edges even at extremely low sampling ratios where conventional baselines fail. By bridging the gap between hardware physics and data-driven processing, this work provides a critical enabler for low-overhead, map-based mode selection in future FAS applications.

\bibliographystyle{IEEEtran}
\bibliography{reference}

\end{document}